\documentclass[epj]{webofc}
\usepackage[varg]{txfonts}   
\wocname{EPJ Web of Conferences}
\woctitle{ICNFP 2016}
\newcommand{\diag}{{\rm \mbox{{  diag}}}}
\newcommand{\ini}{{\rm \mbox{{\scriptsize  ini}}}}

\begin{document}
\selectlanguage{english}
\title{Far-from-equilibrium energy flow and entanglement entropy~\thanks{Plenary talk given by E.~Meg\'{\i}as at the 5th International Conference on New Frontiers in Physics (ICNFP 2016), 6-14 July 2016, Kolymbari, Crete, Greece.}}
%
%

\author{Eugenio Meg\'{\i}as\inst{1,2}\fnsep\thanks{\email{emegias@mppmu.mpg.de}}
}

\institute{Max-Planck-Institut f\"ur Physik (Werner-Heisenberg-Institut), F\"ohringer Ring 6, D-80805, Munich, Germany
\and
          Departamento de F\'{\i}sica Te\'orica, Universidad del Pa\'{\i}s Vasco UPV/EHU, Apartado 644,  48080 Bilbao, Spain
}

\abstract{%
  The time evolution of the energy transport triggered in a strongly coupled system by a temperature gradient is holographically related to the evolution of an asymptotically AdS black brane. We study the far-from-equilibrium properties of such a system by using the AdS/CFT correspondence. In particular, we describe the appearance of a steady state, and study the information flow by computing the time evolution of the holographic entanglement entropy. Some universal properties of the quenching process are presented.
}
\maketitle

\section{Introduction}
\label{intro}

The study of out-of-equilibrium dynamics is a notoriously challenging problem. The hydrodynamic approach has some limitations: it can be applied only for systems close to equilibrium, as it is based on the assumption that the mean free path (time) of particles is much shorter than the characteristic size (time scale) of the system. However, many physical systems are affected by far-from-equilibrium dynamics. Some examples are the the initial stages of the Quark-Gluon plasma thermalization~\cite{Ishii:2015gia}, quenches in some condensed matter systems and fluctuations in the fractional Hall effect~\cite{Polkovnikov:2010yn}. These kinds of studies are very difficult with the current techniques, and nowadays the results are limited to simple models. An interesting and tractable class of non-equilibrium configurations are the steady state flows which emerge between two heat baths~\cite{Bernard:2012je,Chang:2013gba,Bhaseen:2013ypa,Amado:2015uza,Megias:2015tva}. The energy flow between these two systems is characterized by a time-independent configuration, but it does not correspond to equilibrium. 

AdS/CFT is an excellent tool to study the thermalization process of theories with a Conformal Field Theory (CFT) fixed point in the ultraviolet, as it is then mapped to black hole formation in asymptotically AdS space~\cite{Witten:1998zw}. A strong motivation to apply the AdS/CFT correspondence to far-from-equilibrium dynamics is that it might help to establish some universal properties of these systems. In particular, it could give some insight into the universal quenching properties of the systems. Apart from the energy flow, it is very interesting to study the information flow between two systems which are initially isolated. On this respect, some important quantity is the entanglement entropy which measures how the information get exchanged between two systems. Holographically it can be measured as a generalization of the Bekenstein-Hawking entropy formula~\cite{Ryu:2006bv,Ryu:2006ef,Hubeny:2007xt}. Another quantity derived from the entanglement entropy is the mutual information, which measures the amount of information that can be obtained from one of the systems by looking at the other system.

In this work we will study, within the AdS/CFT correspondence, the formation and time evolution of the steady state between two asymptotic equilibrium systems at different temperatures. In a second step we will address the time evolution of the entanglement entropy, and study some universal properties of the quenching process.

\section{Energy flow in strongly coupled systems}
\label{sec:energy_flow}

It was shown in~\cite{Bernard:2012je} the existence of a homogeneous steady state in a class of $(1+1)$-dim CFTs. A universal formula for the heat flow and the generalization to higher dimensions was derived in~\cite{Chang:2013gba,Bhaseen:2013ypa}. We will study in this section the steady state formation in these systems in any number of dimensions.

\subsection{Steady state formation}
\label{subsec:steady_state}

\begin{figure*}[t]
\begin{tabular}{ccc}
\hspace{-0.4cm}\includegraphics[width=63mm]{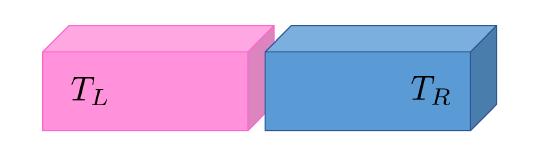} &
\hspace{-0.7cm}\includegraphics[width=20mm]{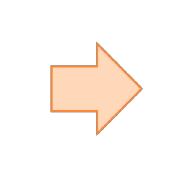} &
\hspace{-0.5cm}\includegraphics[width=63mm]{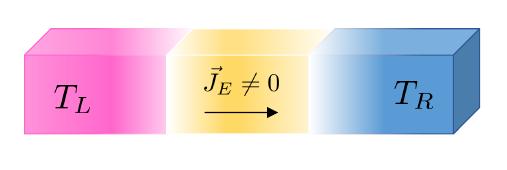} \\
\end{tabular}
\caption{Two isolated systems initially at equilibrium are put in contact at $t=0$. A spatially homogeneous non-equilibrium steady state develops at late times, and it carries an energy current $J_E = \langle T^{tx}\rangle_s$.}
\label{fig:system}
\end{figure*}

Let us consider two thermal reservoirs in $(d+1)$-dim, each of them initially at equilibrium but at different temperatures, $T_L$ and $T_R$. The initial energy density reads
\begin{equation}
\varepsilon(x,t=0) = d a_d \left[ T_L^{d+1} \Theta(-x) + T_R^{d+1} \Theta(x) \right] \,,
\end{equation}
where $a_d$ depends on the number of degrees of freedom in the CFT. After bringing the two systems into thermal contact at $t=0$, a spatially homogeneous steady state develops, carrying a heat flow $J_E$ which transfers energy from the hottest to the coldest system. Such a physical situation is presented in Fig.~\ref{fig:system}. 
The steady state configuration in the CFT can be described by the Lorentz-boosted stress tensor
\begin{equation}
\langle T^{\mu\nu} \rangle = a_d T^{d+1} \left( \eta^{\mu\nu} + (d+1) u^\mu u^\nu \right) \,,   \label{eq:Tboosted}
\end{equation}
where $\eta^{\mu\nu} = \diag(-1,1,\cdots,1)$ is the CFT metric, and $u^\mu = (\cosh\theta, \sinh\theta, 0, \cdots, 0)$ is the fluid velocity with $\theta$ a boost parameter in the $x$-direction~\cite{Bhaseen:2013ypa}. By using the conversation of energy and momentum and traceless of the stress tensor in the CFT,
\begin{equation}
\partial_\mu \langle T^{\mu\nu}\rangle =0  \,, \qquad \langle T^{\mu}_{\mu}\rangle = 0  \,, \label{eq:hydro_eom}
\end{equation}
it has been obtained solutions consisting of ``{\it shockwaves}'' emanating from the interface~\cite{Smoller:1993,Bernard:2012je,Bhaseen:2013ypa,Chang:2013gba,Megias:2015tva}, of the form
\begin{equation}
\langle T^{tt} \rangle = {\cal G}(x - c_s t) + {\cal G}(x + c_st)  \,, \qquad \langle T^{tx} \rangle = {\cal G}(x - c_st) - {\cal G}(x + c_st)    \,, \label{eq:shockwave}
\end{equation}
where ${\cal G}(z)$ is some function which depends on the initial conditions. In this solution the intermediate steady state is bounded by two {\it shocks} propagating outwards at speed $c_s$. In the rest of this section we will study the shockwaves solutions in AdS/CFT in any number of dimensions. In Sec.~\ref{sec:Information_Flow} we will particularize the system to $d=1$, and study the information flow in this case.

\subsection{Holographic model}
\label{subsec:model}

We will present a holographic model to study the system described above at strong coupling . Other out-of-equilibrium stationary configurations in holography can be found in e.g.~\cite{Khlebnikov:2011ka,Fischetti:2012vt,Emparan:2013fha,Ecker:2015kna,Herzog:2016hob} and references therein. Let us consider the Einstein-Hilbert action in $(d+2)$-dim given by
\begin{equation}
S= {1 \over 16\pi G} \int d^{d+2}x \sqrt{-g}\left\{  R + 2\Lambda \right\}  \,, \label{eq:S}
\end{equation}
where $\Lambda = -d(d+1)/2$ is a negative cosmological constant. The equations of motion write
\begin{equation}
R_{MN} - \frac{1}{2}g_{MN} R + g_{MN} \Lambda = 0 \,, \qquad M, N = 1,\cdots, d+2 \,. \label{eq:eom}
\end{equation}
As mentioned above, the energy transport is characterized by a Lorentz-boosted thermal distribution, Eq.~(\ref{eq:Tboosted}). Its gravity dual is a boosted black brane with metric~\cite{Bhaseen:2013ypa}
\begin{equation}
ds^2 = g_{MN} dx^M dx^N = \frac{L^2}{z^2} \left[ \frac{dz^2}{f(z)} 
- f(z) \left( \cosh \theta \, dt - \sinh \theta \, dx \right)^2 
+ \left( \cosh \theta \, dx - \sinh \theta \, dt \right)^2
+ dx_\perp^2
\right] \,, \label{eq:BHsolution}
\end{equation}
where $x_\perp$ parameterizes the transverse spatial coordinates. $L$ is the AdS radius, $f(z) = 1-\left(\frac{z}{z_h}\right)^{d+1}$ and $z_h = \frac{d+1}{4\pi T}$.  It is straightforward to check that Eq.~(\ref{eq:BHsolution}) is a solution of the equations of motion Eq.~(\ref{eq:eom}) as long as the black hole horizon $z_h$ and the boost parameter $\theta$ are constant and uniform, i.e.~$z_h \ne z_h(t,x)$ and $\theta \ne \theta(t,x)$. While this solution is adequate to study the spatially homogeneous steady state regime corresponding to $t \gg |x|$, we expect the existence of more general $(t,x)$-dependent solutions which characterize the formation process of the steady state. In the following we will focus on these solutions.

\subsection{Linearized solution}
\label{sec:linearization}

\begin{figure*}[t]
\begin{tabular}{cc}
\includegraphics[width=6.65cm]{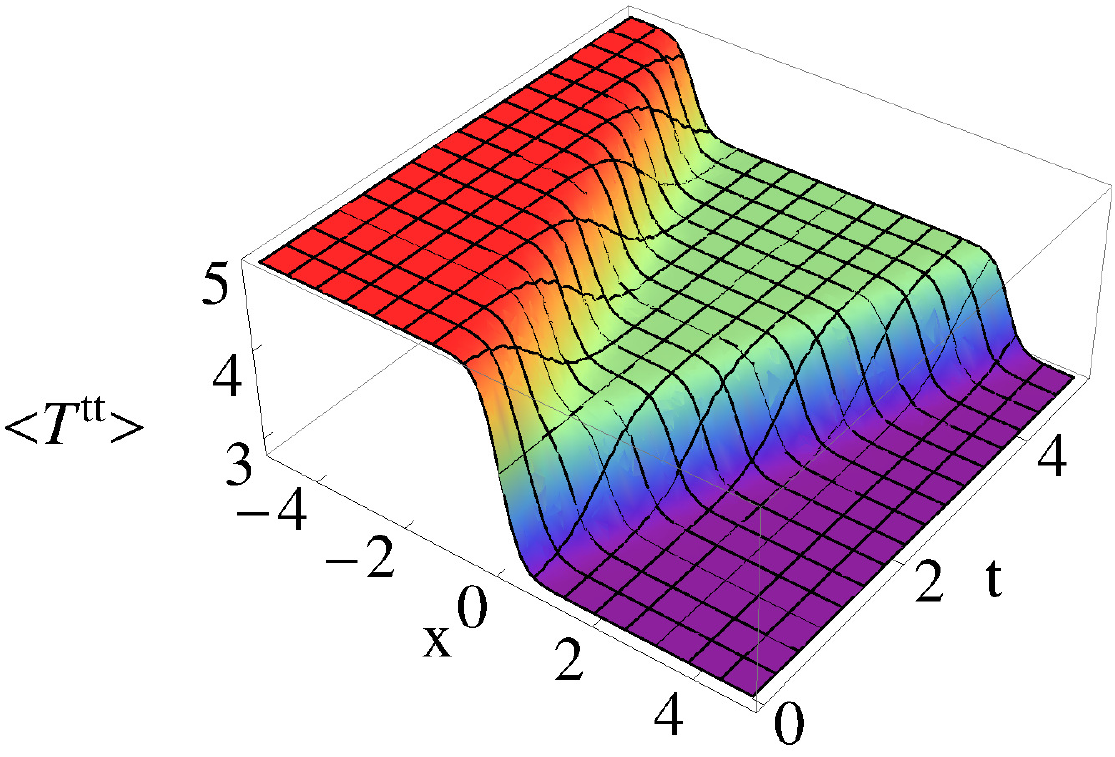} & 
\includegraphics[width=6.65cm]{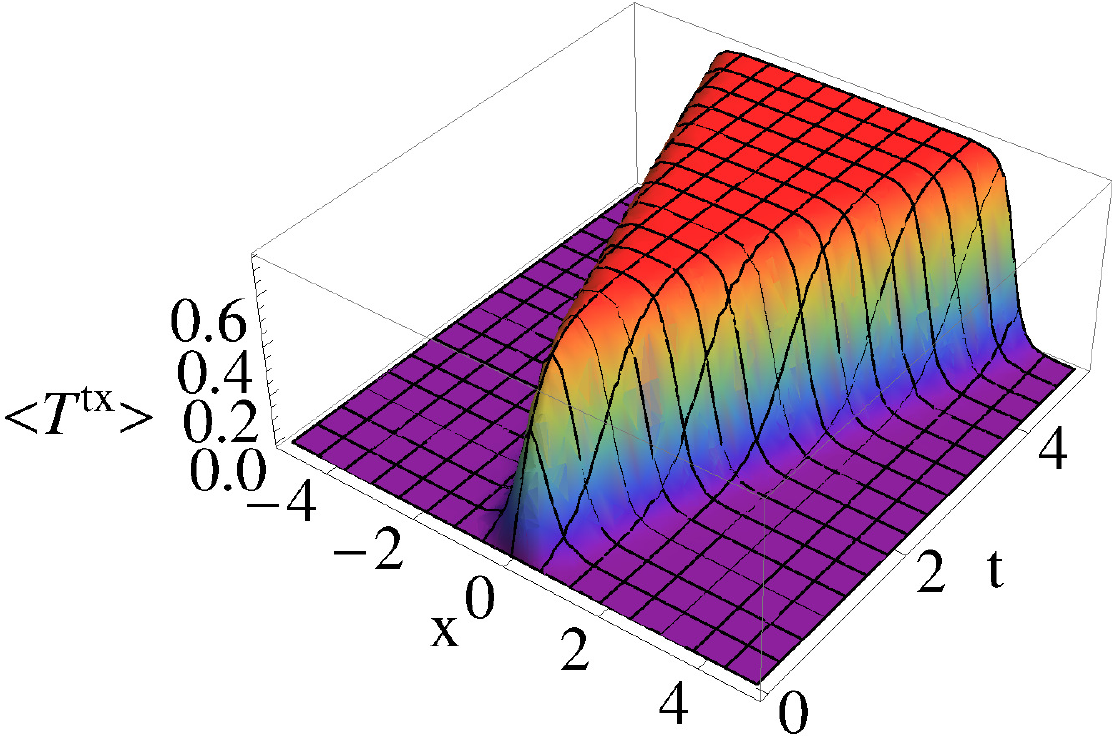}
\end{tabular}
\caption{\it Energy density and energy current computed with the linearized solution given by Eqs.~(\ref{eq:Tttlinearized}) and (\ref{eq:Ttxlinearized}) respectively. It is used the initial profile~$T_{\ini}(x) = \frac{T_R + T_L}{2} + \frac{T_R-T_L}{2}\tanh(\alpha x)$, with $T_L = 1.2$, $T_R=1$ and $\alpha=3$. We have considered $d=2$ and set $G=1$.}
\label{fig:background5}
\end{figure*}

A convenient method to find a solution of Eq.~(\ref{eq:eom}) valid at any point of the space-time consists in the linearization of the problem. Let us consider a solution of the form~\cite{Megias:2015tva}
\begin{equation}
z_h(t,x) = z_{h(0)} + \epsilon \, z_{h(1)}(t,x) + \cdots  \,, \qquad \theta(t,x) = \theta_{(0)} + \epsilon \, \theta_{(1)}(t,x) + \cdots  \,, \label{eq:zhtheta}
\end{equation}
which means that all the space-time dependence is treated as a perturbation around the background $ z_{h(0)}$ and $\theta_{(0)}$, which we keep constant and uniform. Then the equations of motion reduce to the following two equations
\begin{equation}
\partial_t^2 z_{h(1)}(t,x) -  c_s^2 \partial_x^2 z_{h(1)}(t,x) = 0 \,,  \qquad \partial_t z_{h(1)}(t,x)  - c_s^2 z_{h(0)} \partial_x \theta_{(1)}(t,x) = 0   \,,
\end{equation}
where $c_s^2 = 1/d$ is the squared speed of sound. The solution of these equations was obtained in~\cite{Megias:2015tva} under appropriate boundary conditions, in particular
\begin{itemize}
\item $T_{\ini}(x) \equiv T(t=0,x)\,$,
\item $\langle T^{tx}(t=0,x)\rangle = 0 \quad$ for $ \quad x \in (-\infty, +\infty)\,$.
\end{itemize}
The first boundary condition corresponds to the freedom in the initial profile of the temperature~$T_{\ini}(x)$. Only when this profile is non-uniform, the system has a nontrivial time evolution. The second boundary condition states that there is no energy flow at $t=0$, as the two subsystems are initially isolated. Finally, the linearized solution leads to the following $t$-dependent energy density and energy flow~\cite{Megias:2015tva}
\begin{eqnarray}
&&\langle T^{tt}(t,x) \rangle = \frac{d}{8 G} \frac{1}{z_{h(0)}^{d}} \left[ -\frac{d}{2 \pi z_{h(0)}}  + \epsilon \, ( T_\ini(x - c_s t) + T_\ini(x + c_s t) )  \right]   \,,  \label{eq:Tttlinearized} \\
&&\langle T^{tx}(t,x) \rangle = \epsilon \frac{1}{c_s} \frac{1}{8 G } \frac{1}{z_{h(0)}^{d}} \left[ T_\ini(x - c_s t) - T_\ini(x + c_s t) \right]   \,,  \label{eq:Ttxlinearized}
\end{eqnarray}
where $z_{h(0)} = \frac{d+1}{2\pi (T_L + T_R)}$. It is easy to see that corrections of order ${\cal O}(\epsilon)$ are always proportional to factors $\propto \frac{T_L - T_R}{T_L + T_R}$. This illustrates the fact that the $\epsilon$-expansion in Eq.~(\ref{eq:zhtheta}) is equivalent to a small gradient expansion, i.e. $\Big|\frac{T_L - T_R}{T_L + T_R}\Big| \ll 1$, and ultimately to linearized hydrodynamics. This solution leads to the existence of ``{\it shockwaves}'' propagating at speed $c_s$, i.e. it is of the form of Eq.~(\ref{eq:shockwave}) and it fulfills Eq.~(\ref{eq:hydro_eom}).~\footnote{See~\cite{Smoller:1993,Chang:2013gba} for an alternative derivation of this solution by using hydrodynamic considerations.} We display in Fig.~\ref{fig:background5} the numerical result of Eqs.~(\ref{eq:Tttlinearized})-(\ref{eq:Ttxlinearized}). The formation of the steady state and the propagation of the shockwaves is properly described in the regime of small difference of temperatures.

\subsection{Exact solution for $\mathbf{d=1}$}
\label{subsec:d2}

To the best of our knowledge, no exact shockwave solution has been found for the holographic model of Eq.~(\ref{eq:S}) in $d \geq 2$, so that in these cases one has to resort to numerics~\cite{Amado:2015uza} or to linearization methods like the one presented above. However, an exact solution of the model with $d=1$ was found in~\cite{Bhaseen:2013ypa}. This is possible due to the absence of propagating gravitational degrees of freedom. We will present here this solution for completeness. In Fefferman-Graham coordinates it writes
\begin{equation}
ds^2 = \frac{L^2}{z^2} \Big( dz^2 + \tilde{g}_{\mu\nu} dx^\mu dx^\nu \Big) \,, \label{eq:FG}
\end{equation}
where 
\begin{eqnarray}
\tilde{g}_{tt} &=& -\bigg(1-\frac{z^2}{L^2}\left( F(x-t) + F(x+t) \right)\bigg)^2  +  \bigg( \frac{z^2}{L^2}\left( F(x-t) - F(x+t) \right) \bigg)^2  \,,  \\
\tilde{g}_{tx} &=&   -2 \frac{z^2}{L^2}\left( F(x-t) - F(x+t) \right)  \,,  \\
\tilde{g}_{xx} &=& \bigg(1+\frac{z^2}{L^2}\left( F(x-t) + F(x+t) \right)\bigg)^2  -  \bigg( \frac{z^2}{L^2}\left( F(x-t) - F(x+t) \right) \bigg)^2  \,. \label{eq:gxx}
\end{eqnarray}
Note that this solution fulfills the UV boundary condition~$\tilde{g}_{\mu\nu}(z=0) = \eta_{\mu\nu}$. The vacuum expectation value of the energy-momentum tensor writes
\begin{eqnarray}
\langle T^{tt} \rangle \;\;\; = \;\;\; \langle T^{xx}\rangle  &=&  \frac{c}{6\pi L^2} \left( F(x-t) + F(x+t) \right)   \,, \\
\langle T^{tx} \rangle &=& \frac{c}{6\pi L^2} \left( F(x-t) - F(x+t) \right) \,, 
\end{eqnarray}
where $c$ is the central charge of the CFT, and it fulfills the initial condition $\langle T^{tx}(t=0,x)\rangle = 0$. For numerical computations we will consider the profile
\begin{equation}
F(v) = \frac{\pi^2 L^2}{4} \Big( (T_L^2 + T_R^2) + (T_R^2 - T_L^2) \tanh(\alpha v) \Big) \,. \label{eq:F}
\end{equation}
This function tends to a stepwise function $F(v) \to \frac{\pi^2 L^2}{2} \left( T_L^2 \Theta(-v) + T_R^2 \Theta(v) \right)$ in the limit $\alpha \to \infty$. We will use this solution in the next section.

\section{Information flow}
\label{sec:Information_Flow}

In the previous section we have studied the problem of energy flow between two initially isolated systems. The question then arises: how does information get exchanged between these two systems? Our next goal is to answer this question, and to this end we will study the information flow in the system.

\subsection{Entanglement entropy}
\label{subsec:Entanglement_Entropy}

On the field theory side the {\it entanglement entropy} is defined in the following way. Let us divide the system into two subsystems $A$ and $B$. Then the entanglement entropy of the subsystem~$A$ is defined as the von Neumann entropy of the reduced density matrix obtained by tracing out the degrees of freedom of the subsystem~$B$. On the gravity side, the entanglement entropy can be computed as a generalization of the Bekenstein-Hawking entropy formula~\cite{Ryu:2006bv,Ryu:2006ef,Hubeny:2007xt}. It follows from the area law relation
\begin{equation}
{\cal S}_A = \frac{\textrm{Area}(\gamma_A)}{4 G} \,,
\end{equation}
where the manifold $\gamma_A$ is the $d$-dimensional minimal surface in AdS$_{d+2}$ whose boundary is given by $\partial A$, and $G$ is the $(d+2)$-dimensional Newton constant. The computation of the holographic entanglement entropy requires the study of minimal surfaces, and this demands the solution of geodesic equations. 

In the following we will restrict to the case $d=1$ for simplicity, with coordinates $(t,x,z)$. We will study minimal surfaces whose boundary at $z=0$ is in $x = x_L$ and $x = x_R$, and consider space-like intervals with $t(x_L) = t(x_R)$. The minimal surface compatible with these boundary conditions corresponds to a geodesic line in AdS$_3$, and it follows from a solution of the geodesic equations, which read
\begin{equation}
\frac{d^2 x^P}{ds^2}  +\Gamma^P_{MN} \frac{dx^M}{ds} \frac{dx^N}{ds} = 0 \,, \qquad P= t,x,z \,.  \label{eq:geodesics}
\end{equation}
The induced metric on the minimal surface reads
\begin{equation}
h_{ab} = \frac{\partial x^M}{\partial x^a} \frac{\partial x^N}{\partial x^b} g_{MN} = h_{ss}  \,,
\end{equation} 
where $s$ is the coordinate of the surface. The entanglement entropy then follows from the area of the manifold $\gamma_A$, which can be computed from the induced metric as
\begin{equation}
{\cal S}_A = \frac{1}{4 G} \int_{s(x_L)}^{s(x_R)} ds \, {\cal L} \,, \qquad \textrm{with} \qquad {\cal L} = \sqrt{h_{ss}} \,. \label{eq:S1}
\end{equation}
At this point it is convenient to consider $s$ as an affine parameter, so that
\begin{equation}
\frac{\partial x^M}{\partial s}\frac{\partial x^N}{\partial s} g_{MN}   = 1 \,,
\end{equation}
and the entanglement entropy of Eq.~(\ref{eq:S1}) reduces to the trivial integration ${\cal S}_A = \frac{1}{4 G} \int_{s(x_L)}^{s(x_R)} ds$. The solution of the geodesic equations leads to the behavior $z \sim e^{-|s|}$ in the regime $s \to \pm \infty$, and then one concludes that the entanglement entropy is divergent. In the present case the divergence behaves as~${\textrm{Area}}(\gamma_A^{div}) \sim -2L \log z_{uv} + \cdots$, and a renormalization scheme is required. We will use a minimal subtraction scheme, so that the renormalized entanglement entropy is defined as
\begin{equation}
{\cal S}_A^{ren} = \frac{1}{4 G} \left( \textrm{Area}(\gamma_A) -  \textrm{Area}(\gamma_A^{div}) \right)  \qquad \textrm{with} \qquad   \textrm{Area}(\gamma_A^{div}) = -2 L \log z_{uv} \,. \label{eq:Sreg}
\end{equation}
In the rest of the manuscript we will compute renormalized entropies according to this formula, and we will omit the superscript~$ren$.

\subsection{Geodesic equations}
\label{subsec:geodesic_equations}

The geodesic equations of Eq.~(\ref{eq:geodesics}) consist of three coupled differential equations of second order, whose solution can be expressed in the parametric form
\begin{equation}
t= t(s) \,, \qquad x = x(s) \,, \qquad z=z(s) \,. 
\end{equation}
These equations can be solved by imposing six boundary conditions, which are:
\begin{equation}
\begin{cases}
t(s_L) = t(s_R) = t_0       &   \\
x(s_L) = x_L \,, \qquad x(s_R) = x_R    &   \\
z(s_L) = z(s_R) = z_{uv}      &  
\end{cases} \label{eq:bc}
\end{equation}
We will use the shooting method for the numerical computation of the geodesic equations: we shoot the solution from $s=0$ with given values of $\{t(0),x(0),z(0)\}$ and $\{t^\prime(0),x^\prime(0),z^\prime(0)\}$, and then find the values of these initial conditions that lead to the desired boundary values at $s \to s_{L,R}$.~\footnote{There are in the literature other numerical methods for the solution of this two-point boundary value problem. An example are the relaxation methods, in which the solution is determined by starting with an initial guess and improving it iteratively, see e.g.~\cite{Ecker:2015kna}.}

\begin{figure*}[t]
\includegraphics[width=6.65cm]{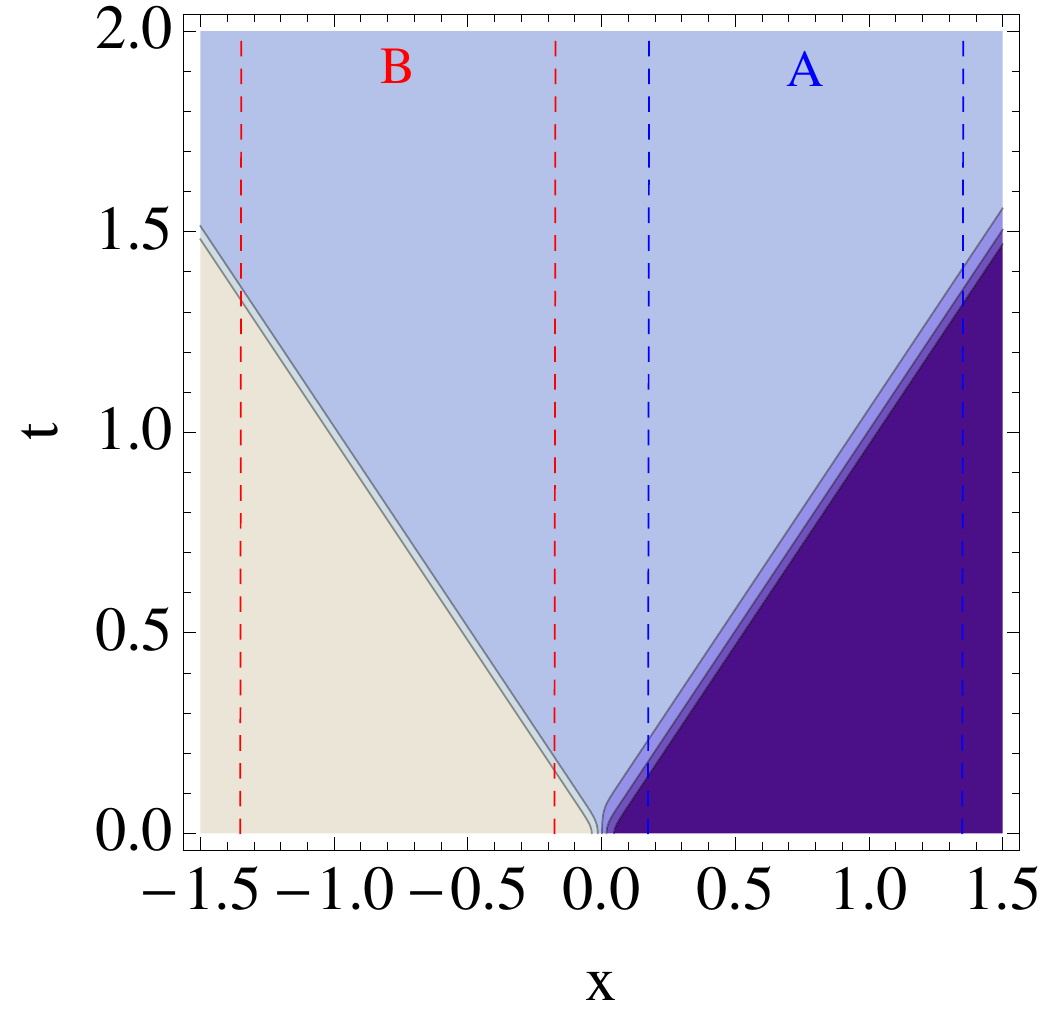} \hspace{2pc}%
\begin{minipage}{14.8pc}
\vspace{-4cm}\caption{\it Contour plot of energy density $\langle T^{tt}(t,x)\rangle$ with the model in $d=1$, cf. Sec.~\ref{subsec:d2}. Dashed lines are the time evolution of the extrema of the intervals $A$ and $B$, in the positive and negative semiplane respectively.  We consider the intervals $x^A \in [0.175,1.35]$ (blue) and $x^B \in [-1.35,-0.175]$ (red), temperatures $T_L = 0.2$, $T_R=0.195$ and $\alpha = 25$.}
\label{fig:Tttd1}
\end{minipage}
\end{figure*}

\begin{figure*}[t]
\begin{tabular}{cc}
\includegraphics[width=7.0cm]{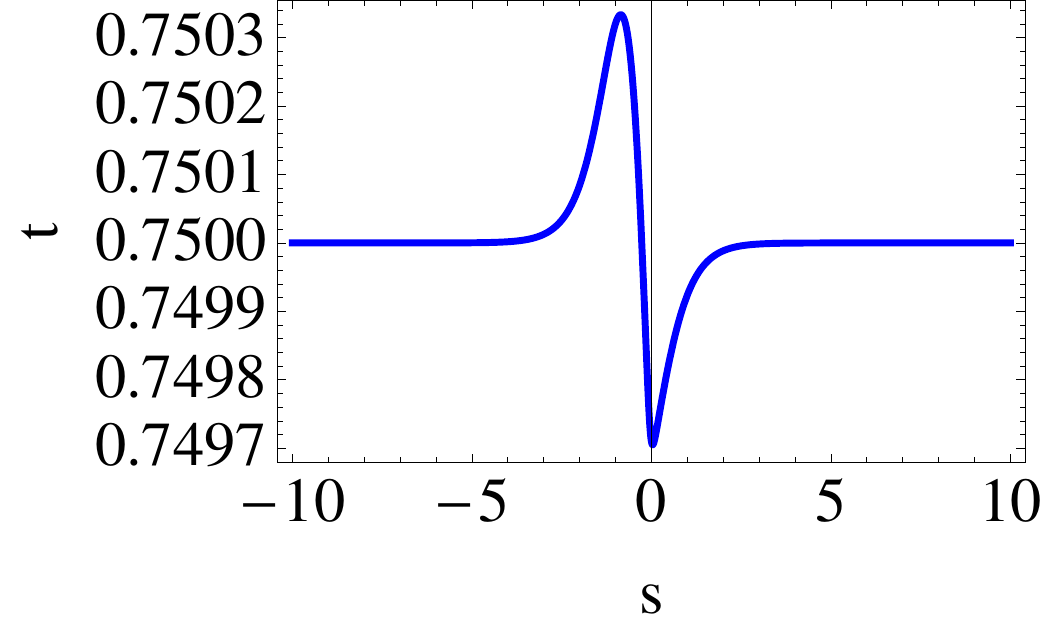} & 
\includegraphics[width=6.35cm]{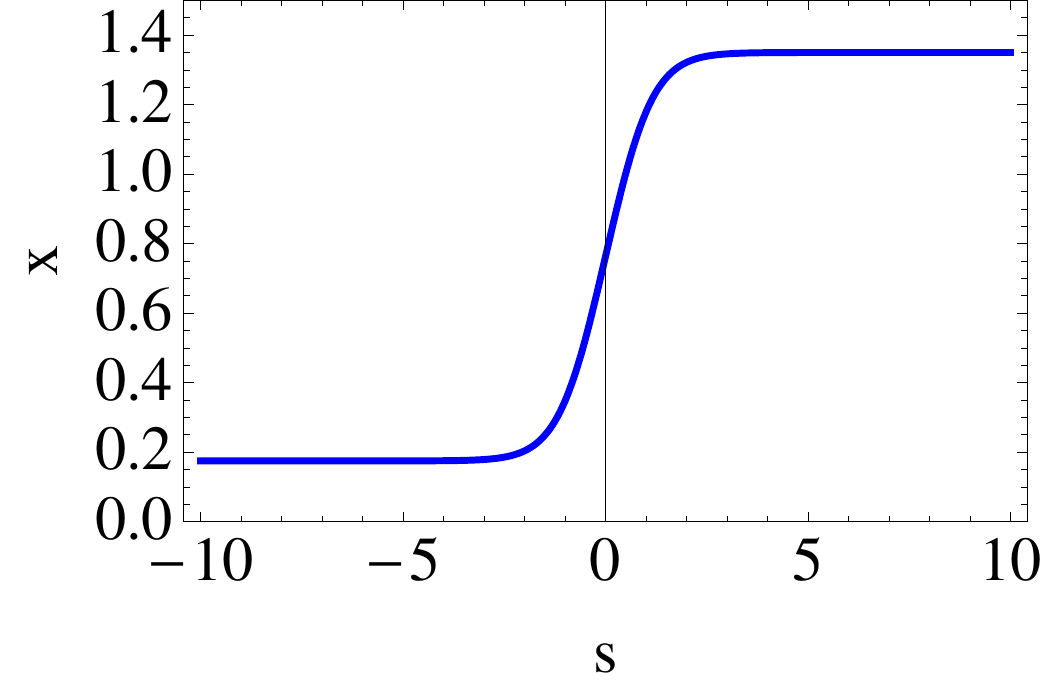} \\
\end{tabular}
\caption{\it Parametric dependence of the geodesic as a function of the affine parameter~$s$. We show $t = t(s)$ (left) and $x = x(s)$ (right). We have considered the interval $x^A \in [0.175,1.35]$ as shown in Fig.~\ref{fig:Tttd1}, and $t_0 = 0.75$, cf. Eq.~(\ref{eq:bc}).}
\label{fig:geodesic1}
\end{figure*}

\begin{figure*}[htb]
\begin{tabular}{cc}
\includegraphics[width=6.65cm]{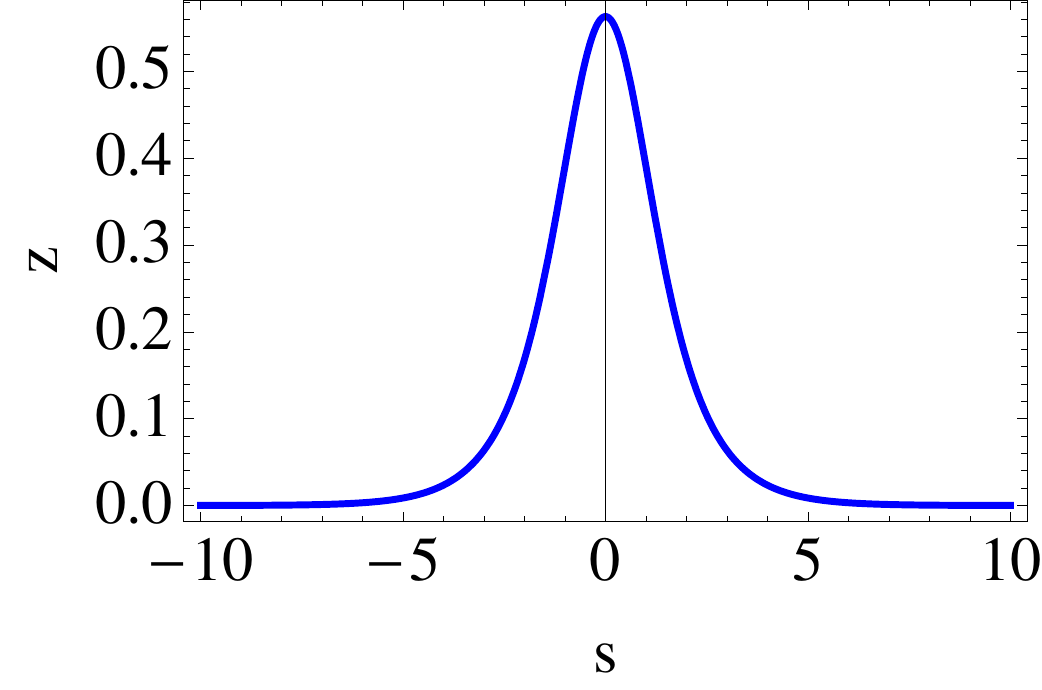} & 
\includegraphics[width=5.65cm]{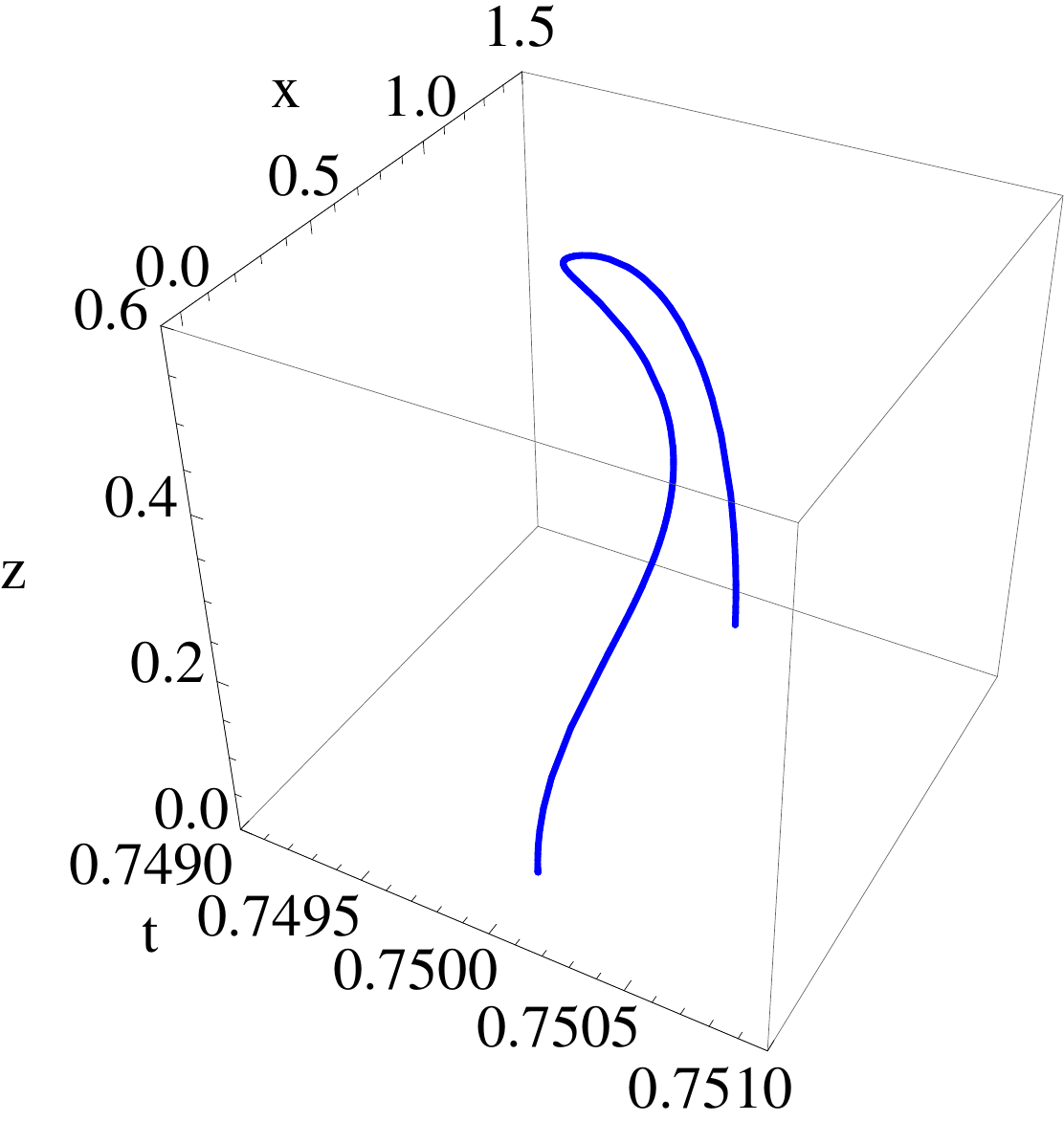}
\end{tabular}
\caption{\it  Parametric dependence of the geodesic. (Left) We show $z = z(s)$. (Right) Geodesic in the space $(t,x,z)$. See Fig.~\ref{fig:geodesic1} for further details.}
\label{fig:geodesic2}
\end{figure*}

We have introduced a cutoff $z_{uv} \ll 1$ to regularize the problem, and this induces also a cutoff in the affine parameter, i.e. $s_L \sim -|\log z_{uv}|$ and $s_R \sim |\log z_{uv}|$. In the following we will consider space-like intervals $A$ and $B$ as shown in Fig.~\ref{fig:Tttd1}. It is displayed in Figs.~\ref{fig:geodesic1} and~\ref{fig:geodesic2} a typical solution of the geodesic equations, which fulfills the boundary conditions of Eq.~(\ref{eq:bc}). Once the geodesics are obtained, the next step is to compute the area of these curves and then the entanglement entropy from Eq.~(\ref{eq:Sreg}). In the rest of this section we will present some results for the time evolution of the entanglement entropy in the system of Sec.~\ref{sec:energy_flow}.

\subsection{Entanglement entropy and universal time evolution}
\label{subsec:entanglement_entropy}

\begin{figure*}[htb]
\begin{tabular}{cc}
\includegraphics[width=6.85cm]{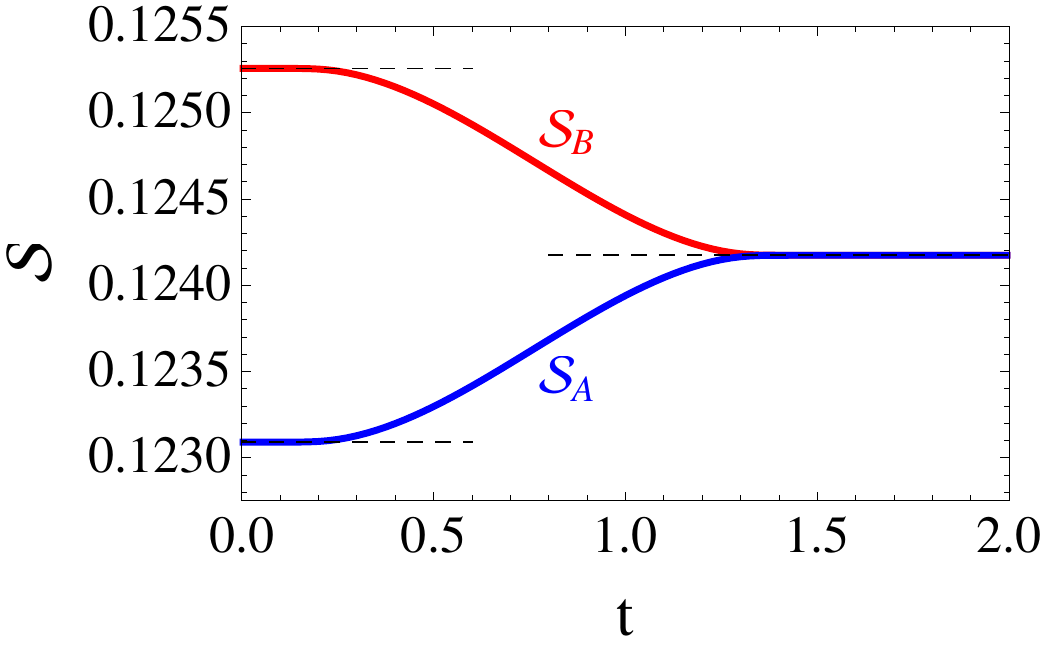} & 
\includegraphics[width=6.45cm]{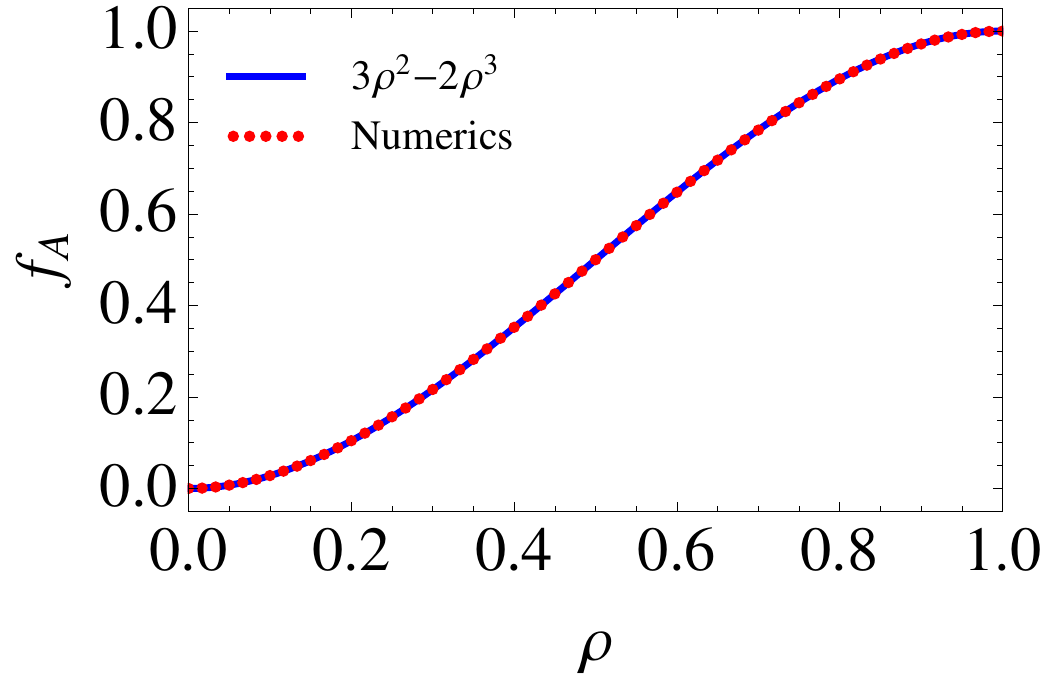}
\end{tabular}
\caption{\it (Left) Entanglement entropies of intervals $A$ and $B$ as a function of time, see Fig.~\ref{fig:Tttd1}. The (dashed) horizontal lines correspond to the results by using the analytical formulas, Eqs.~(\ref{eq:St0}) and (\ref{eq:Sboost}). We have set $G=1$ and $L=1$. (Right) Entanglement entropy ${\cal S}_A$ as a function of time, normalized to $[0,1]$ in both horizontal and vertical axes, cf. Eq.~(\ref{eq:fA}). The dots correspond to the numerical result with the interval $A$ in Fig.~\ref{fig:Tttd1}, while the continuous line is the universal behavior~$f_A(\rho) = 3\rho^2 - 2\rho^3$. }
\label{fig:Suniversal}
\end{figure*}

For the moment we will consider a single interval $x \in [x_L,x_R]$, that we denote by $A$, placed in the positive semiplane, i.e. $x_{L,R}>0$. Let us study the time evolution of the entanglement entropy ${\cal S}_A$ during the process of quenching. We are considering the model with $d=1$, so that the shockwaves are at $t = |x|$. This means that the shockwaves touch the two ends of the interval at times $t=|x_L|$ and $t=|x_R|$, see Fig.~\ref{fig:Tttd1}. We will denote these values as $t_L$ and $t_R$ respectively. If one assumes the limit $\alpha \to \infty$ in Eq.~(\ref{eq:F}), the entanglement entropy turns out to be constant in the regimes $0 \le t \le t_L$ and $t_R \le t$, and there is a non trivial time evolution only in the interval $t_L \le t \le t_R$, i.e.
\begin{equation}
{\cal S}_A(t) =
\begin{cases}
{\cal S}_A(t=0)       &   0 \le t \le t_L \\
{\cal S}_A(t)         &   t_L  \le   t  \le t_R  \\
{\cal S}_A(t=\infty)  &   t_R \le t  
\end{cases}  \,.  \label{eq:SAt}
\end{equation} 
We display in Fig.~\ref{fig:Suniversal} (left) the time evolution of the entanglement entropy of interval $A$ of Fig.~\ref{fig:Tttd1}, from a numerical computation of the geodesic equations. Let us focus on the regime $t_L  \le   t  \le t_R $. It is convenient to define the normalized entanglement entropy~$f_A(\rho)$ as
\begin{equation}
f_A(\rho ) \equiv \frac{{\cal S}_A(t) - {\cal S}_A(t=0)}{{\cal S}_A(t=\infty) - {\cal S}_A(t=0)} \qquad \textrm{with} \qquad \rho \equiv (t-t_0)/\Delta t  \,, \label{eq:fA} 
\end{equation}
where $t_0 = t_L$ and $\Delta t = \ell = |x_R - x_L|$. This corresponds to the function ${\cal S}_A(t)$ normalized to the interval [0,1] in both horizontal and vertical axes. It is clear from Eqs.~(\ref{eq:SAt}) and (\ref{eq:fA}) that $f_A(\rho)$ has the extreme values $f_A(0)=0$ and $f_A(1)=1$. We have computed numerically the entanglement entropy ${\cal S}_A(t)$ in a wide variety of configurations with different temperatures $T_L$, $T_R$ and lengths~$\ell$, and find that the behavior of $f_A(\rho)$ in every case is
\begin{equation}
f_A(\rho ) \simeq 3\rho^2-2\rho^3  \,, \qquad 0 \le \rho \le 1 \,. \label{eq:fAuniversal}
\end{equation}
This function fits extremely well the numerical results of the entanglement entropies up to moderate gradients, i.e. $|T_L - T_R| < T_L + T_R$. This is illustrated in Fig.~\ref{fig:Suniversal} (right) for a particular case. The result of Eq.~(\ref{eq:fAuniversal}) is independent of the values of the parameters $T_L$, $T_R$ and $\ell$, and so it implies the existence of an {\it 'almost'} universal time-evolution of entanglement entropy in the theory with $d=1$.~\footnote{The deviations from universality are stronger for bigger temperature differences.} 

The analysis presented above applies also to intervals in the negative semiplane. We show in Fig.~\ref{fig:Suniversal}~(left) the entanglement entropy of interval $B$ of Fig.~\ref{fig:Tttd1}.  Note that both functions, ${\cal S}_A(t)$ and ${\cal S}_B(t)$, tend to the same value when the intervals reach the steady state regime.

\subsection{Conservation of entanglement entropy}
\label{sec:conservation_entanglement_entropy}

Let us consider the two extreme regimes $t=0$ and $t \to \infty$. It is possible to obtain analytical results for the entanglement entropies in these cases for the model with $d=1$ presented in Sec.~\ref{subsec:d2}. When $t=0$ the metric corresponds to a stepwise black hole, i.e.
\begin{equation}
ds^2 = ds_L^2 \Theta(-x) + ds_R^2 \Theta(x) \,.
\end{equation}
If we place the interval in just one semiplane, i.e. $x_{L,R} > 0$ (or $x_{L,R} < 0$), the entanglement entropy corresponds to the one for a stationary black hole at temperature~$T$, which reads
\begin{equation}
{\cal S}(T,\ell;t=0) = \frac{L}{2G} \log \left( \frac{1}{\pi T}  \sinh( \pi \ell T) \right) \,, \qquad \ell := |x_R - x_L| \,.  \label{eq:St0}
\end{equation}
In this equation $T=T_{L}$ (or $T_R$) when $x_{L,R}<0$ (or $x_{L,R}>0$).  In the other extreme, $t \to \infty$, the system is in the steady state regime, and the entanglement entropy is the one for a boosted black hole, that reads~(see~\cite{workinprogress} for a derivation)~\footnote{Note that Eq.~(\ref{eq:Sboost}) is valid when $t \ge \max(|x_L|,|x_R|)$ if the initial profile $F(v)$ in Eq.~(\ref{eq:F}) is a stepwise function, i.e. in the limit $\alpha \to \infty$. When~$F(v)$ is a smooth function, the rhs of Eq.~(\ref{eq:Sboost}) corresponds to the asymptotic value of the entanglement entropy at very late times, i.e. for $t \gg \max(|x_L|,|x_R|)$.}
\begin{equation}
{\cal S}(T_L,T_R,\ell;t=\infty) =  \frac{L}{4G} \log \left( \frac{1}{\pi^2 T_L T_R} \sinh\left( \pi \ell T_L \right)  \sinh\left( \pi \ell  T_R \right) \right) \,. \label{eq:Sboost}
\end{equation}
These analytical results, Eqs.~(\ref{eq:St0}) and (\ref{eq:Sboost}), correspond to ${\cal S}_A(t=0)$ and ${\cal S}_A(t=\infty)$ in Eq.~(\ref{eq:SAt}), respectively. From these formulas one can easily obtain the following property:
\begin{equation}
{\cal S}_A(t=0) + {\cal S}_B(t=0) = {\cal S}_A(t=\infty) + {\cal S}_B(t=\infty) \,, \label{eq:SASBconservation}
\end{equation}
where we have considered intervals $A$ with $x^A_{L,R}>0$, and $B$ with $x^B_{L,R}<0$, and lengths $\ell = \ell_A = \ell_B$. This property is non trivial, as in the lhs of Eq.~(\ref{eq:SASBconservation}) there is the contribution of stationary black hole solutions at temperatures $T_L$ and $T_R$, while in the rhs there is a boosted black hole and the corresponding energy flow contributes as well to the entanglement entropy. This relation is very interesting as it implies the {\it 'conservation'} of entanglement entropies between $t=0$ and $t=\infty$. 

\begin{figure*}[t]
\begin{tabular}{cc}
\includegraphics[width=7.15cm]{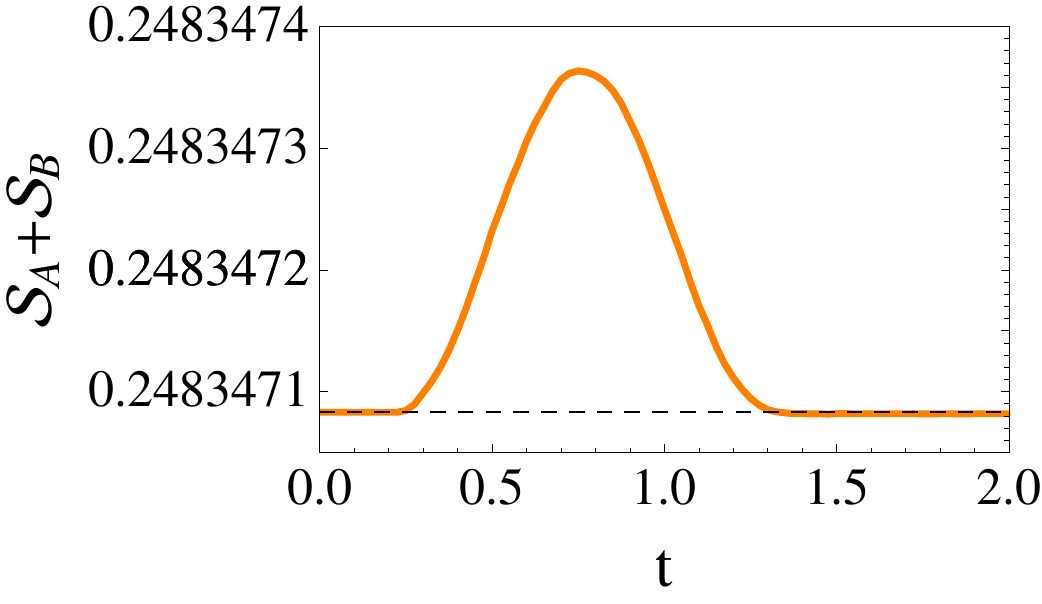} & 
\includegraphics[width=6.15cm]{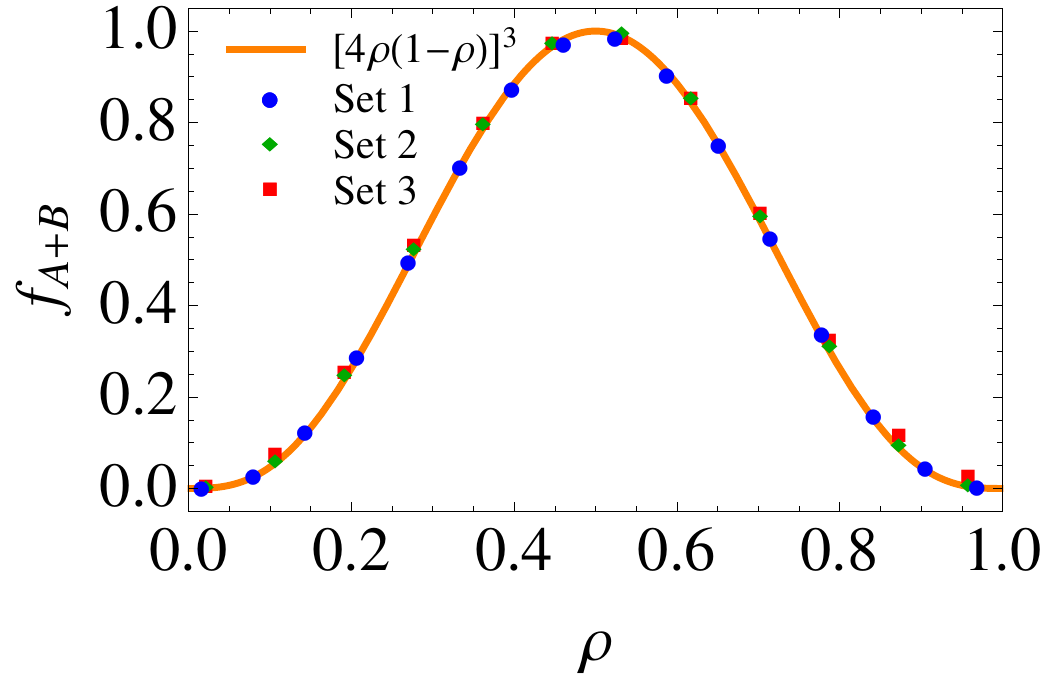}
\end{tabular}
\caption{\it (Left) Entanglement entropy ${\cal S}_A + {\cal S}_B$ as a function of time, see Fig.~\ref{fig:Tttd1}.  The (dashed) horizontal line corresponds to the result by using the analytical formulas, Eqs.~(\ref{eq:St0}) and (\ref{eq:Sboost}). We have set $G=1$ and $L=1$. (Right) Entanglement entropy ${\cal S}_{A+B}$ as a function of time, normalized to $[0,1]$ in both horizontal and vertical axes, cf. Eq.~(\ref{eq:fAB}). The dots correspond to the numerical result with intervals $A$ and $B$, placed symmetrically with respect to $x=0$ as shown in Fig.~\ref{fig:Tttd1}, in different configurations: Set 1 is $(T_L=0.2, T_R=0.195, \ell_A = \ell_B =  1.175)$, Set 2 is $(T_L=0.2, T_R=0.175,  \ell_A = \ell_B  = 1.175)$ and Set 3 is $(T_L=0.2, T_R=0.175,  \ell_A = \ell_B = 1.475)$. The continuous line is the universal behavior~$f_{A+B}(\rho) = \left[ 4\rho(1-\rho)\right]^3 $.}
\label{fig:SASBpeak}
\end{figure*}

It is displayed in Fig.~\ref{fig:SASBpeak} (left) the time evolution of ${\cal S}_{A+B} \equiv {\cal S}_A + {\cal S}_B$. We can see that our numerics confirms the conservation rule of Eq.~(\ref{eq:SASBconservation}). In the next subsection we will study this system in the quenching regime, i.e. $t_L \le t \le t_R$ in Eq.~(\ref{eq:SAt}), and characterize some violations of the entanglement entropy conservation in this case. 

\subsection{Non-universal effects in time evolution}
\label{subsec:non_universal}

As it is shown in Fig.~\ref{fig:SASBpeak} (left), we find from our numerics that ${\cal S}_{A+B}(t) \ne \textrm{cte}$ in the quenching regime. This implies that the entanglement entropy is not exactly conserved at intermediate times. A straightforward computation shows that these non-conservation effects are only possible if there are non-universal contributions in Eq.~(\ref{eq:fAuniversal}), otherwise this equation would predict ${\cal S}_{A+B}(t) = \textrm{cte}$.

In the following we will restrict to intervals $A$ and $B$ with the same length and placed symmetrically with respect to $x=0$, i.e. $\ell_A = \ell_B$ and $x_{L,R}^A = -x_{R,L}^B$. While the function ${\cal S}_{A+B}(t)$ has the same value at $t=0$ and $t=\infty$ (see Eq.~(\ref{eq:SASBconservation})), we find from our numerics that it has a maximum at $t_{\textrm{max}} = (t_L + t_R)/2$. In order to characterize the time evolution of ${\cal S}_{A+B}(t)$, let us define the normalized entanglement entropy
\begin{equation}
f_{A+B}(\rho) \equiv \frac{{\cal S}_{A+B}\left( t \right)-{\cal S}_{A+B}(t=0)}{{\cal S}_{A+B}(t_{\textrm{max}})-{\cal S}_{A+B}(t=0)} \qquad \textrm{with} \qquad \rho \equiv (t-t_0)/\Delta t  \,,  \label{eq:fAB}
\end{equation}
where $t_0$ and $\Delta t$ are defined as in Eq.~(\ref{eq:fA}). Finally, from a numerical computation of~$f_{A+B}(\rho)$ in a wide variety of intervals, we find that its behavior is universal and it is given by
\begin{equation}
f_{A+B}(\rho) \simeq \left[ 4\rho(1-\rho)\right]^3 \,, \qquad 0 \le \rho \le 1 \,. \label{eq:fABuniversal}
\end{equation}
This is illustrated in Fig.~\ref{fig:SASBpeak} (right) for several configurations. From a combination of the results in Eqs.~(\ref{eq:fAuniversal}) and (\ref{eq:fABuniversal}), we conclude that the normalized entanglement entropy defined in Eq.~(\ref{eq:fA}) can be written as
\begin{equation}
f_A(\rho) = 3 \rho^2 - 2 \rho^3 + \Delta_A(\rho)   \,, \quad  \textrm{with} \quad \Delta_A(\rho) \simeq   C(T_L, T_R, \ell)  \cdot [4\rho(1-\rho)]^3 \,. \label{eq:fA2}
\end{equation}
The factor $C(T_L, T_R, \ell)$ has a non-universal dependence on the parameters of the interval, so that $\Delta_A(\rho)$ is a non-universal contribution to $f_A(\rho)$. Note, however, that $C(T_L,T_R,\ell)$ does not affect the universal behavior of $f_{A+B}(\rho)$, see Eq.~(\ref{eq:fABuniversal}). Some remarks deserve to be mentioned: on the one hand, $\Delta_A(\rho)$ is a correction of order ${\cal O}(\rho^3)$, so that it does not jeopardize the behavior ${\cal S}_A(t) \sim t^2$ which seems to be general in a wide variety of systems. On the other hand, the effect of $\Delta_A(\rho)$ is extremely small in the configurations we have studied numerically.~\footnote{One can see from Fig.~\ref{fig:SASBpeak} (left) that in this case the peak in ${\cal S}_{A+B}(t_{\textrm{max}})$ is a correction of order ${\cal O}(10^{-6})$ with respect to ${\cal S}_{A+B}(0)$, so that the order of magnitude of the non-universal contribution in Eq.~(\ref{eq:fA2}) is 
\begin{equation}
C(T_L,T_R,\ell) \simeq \Delta_A\left(\rho=\frac{1}{2}\right)  \simeq \frac{1}{2} \frac{ {\cal S}_{A+B}(t_{\textrm{max}}) - {\cal S}_{A+B}(0)  }{  {\cal S}_A(\infty)-{\cal S}_A(0) } \sim {\cal O}(10^{-4}) \,.
\end{equation}
}

\section{Conclusions and outlook}

In this work we have studied a holographic model for far-from-equilibrium dynamics that allows to describe the time dependent properties of energy flow and information flow of two thermal reservoirs initially isolated. Of relevance in this system is the appearance of a universal steady state, described by a boosted black brane. By using a linearization method, we have described the formation of the steady state and the propagation of shockwaves for any space-time dimension. A relevant observable that provides physical insight into the evolution of the system is the entanglement entropy, which measures the information flow between two subsystems. By using the exact solution for $d=1$ provided in~\cite{Bhaseen:2013ypa}, we have studied numerically the time evolution of the entanglement entropy, and characterized some universal properties of the quenching process.

There remain some open questions. It would be interesting to study the entanglement entropy at higher dimensions, i.e. for $d \ge 2$. This can be easily done by using the linearized background presented in this work. An analysis beyond the linear response regime, i.e. for~$0 < T_R/T_L <1$, would demand a full numerical solution of the equations of motion for the background, see e.g.~\cite{AbajoArrastia:2010yt,Chesler:2013lia,Amado:2015uza,Ecker:2015kna,workinprogress}. A quantity of interest related to the entanglement entropy is the {\it mutual information}, which measures which information of subsystem $A$ is contained in subsystem $B$. An advantage of this quantity is that it is finite, so that it does not need to be regularized. The Ryu-Takayanagi formula for entanglement entropy satisfies some inequalities that set special restrictions, and have profound implications for the structure of holographic quantum states and their correlations. Apart from the monogamy of mutual information and strong subadditivity, other inequalities involving a large number of subsystems have been proved in the static case, see e.g.~\cite{Bao:2015bfa}. It would be interesting the study the time evolution of the mutual information, and to check the validity of the entanglement entropy inequalities in time dependent systems. These and other issues will be addressed in a forthcoming publication~\cite{workinprogress}.

%
%
%

\begin{acknowledgement}
I would like to thank J.~Erdmenger, D.~Fern\'andez, M.~Flory, A.K.~Straub and P.~Witkowski for valuable discussions. This work has been supported by Plan Nacional de Altas Energ\'{\i}as Spanish MINECO grant FPA2015-64041-C2-1-P, and by the Spanish Consolider Ingenio 2010 Programme CPAN (CSD2007-00042). I thank the Instituto de F\'{\i}sica Te\'orica UAM/CSIC, Madrid, Spain, for their hospitality during the completion of the final stages of this work. The research of E.M. is supported by the European Union under a Marie Curie Intra-European fellowship (FP7-PEOPLE-2013-IEF) with project number PIEF-GA-2013-623006, and by the Universidad del Pa\'{\i}s Vasco UPV/EHU, Bilbao, Spain, as a Visiting Professor.
\end{acknowledgement}


\end{document}